\documentclass[11pt]{article}

\usepackage{amsfonts,amsmath,amssymb}
\usepackage{pictex,color}
\usepackage{boxedminipage}
\usepackage{epsfig}

\newcommand{\ket}[1]{| #1 \rangle}

\renewcommand{\phi}{\varphi}

\newcommand{\floor}[1]{\left\lfloor #1 \right\rfloor}
\newcommand{\ceil}[1]{\left\lceil #1 \right\rceil}

\newcommand{\GF}{\textnormal{GF}}
\renewcommand{\O}{\mathcal{O}}
\renewcommand{\iff}{\hspace{2mm}\leftrightarrow\hspace{2mm}}

\newcommand{\qed}{\hspace{5mm}\square\vspace{4mm}}
\newcommand{\proof}{\textbf{Proof: }}
\renewcommand{\mod}[1]{\text{mod } #1}

\newcommand{\XOR}{\textnormal{XOR}}
\newcommand{\SWAP}{\textnormal{SWAP}}

\renewcommand{\,}{\hspace{1mm},\hspace{1mm}}

\newtheorem{claim}{Claim}
\newtheorem{cor}{Corollary}

\begin{document}
\title{\textbf{Optimized quantum implementation of elliptic curve arithmetic over binary fields}}
\author{Phillip Kaye \thanks{prkaye@iqc.ca,
School of Computer Science, University of Waterloo, Waterloo, ON,
Canada.} and Christof Zalka
\thanks{zalka@iqc.ca, Department of Physics, University of
Waterloo.}}

\maketitle

\begin{abstract}
Shor's quantum algorithm for discrete logarithms applied to elliptic
curve groups forms the basis of a ``quantum attack'' of elliptic
curve cryptosystems.  To implement this algorithm on a quantum
computer requires the efficient implementation of the elliptic curve
group operation.  Such an implementation requires we be able to
compute inverses in the underlying field.  In \cite{PZ03}, Proos and
Zalka show how to implement the extended Euclidean algorithm to
compute inverses in the prime field $\GF(p)$.  They employ a number
of optimizations to achieve a running time of $O(n^2)$, and a
space-requirement of $O(n)$ qubits (there are some trade-offs that
they make, sacrificing a few extra qubits to reduce running-time).
In practice, elliptic curve cryptosystems often use curves over the
binary field $\GF(2^m)$. In this paper, we show how to implement the
extended Euclidean algorithm for polynomials to compute inverses in
$\GF(2^m)$. Working under the assumption that qubits will be an
`expensive' resource in realistic implementations, we optimize
specifically to reduce the qubit space requirement, while keeping
the running-time polynomial.  Our implementation here differs from
that in $\cite{PZ03}$ for $\GF(p)$, and we are able to take
advantage of some properties of the binary field $\GF(2^m)$. We also
optimize the overall qubit space requirement for computing the group
operation for elliptic curves over $\GF(2^m)$ by decomposing the
group operation to make it ``piecewise reversible'' (similar to what
is done in \cite{PZ03} for curves over $\GF(p)$).
\end{abstract}

\section{Introduction}

A very significant potential application of quantum computers lies
in their ability to efficiently solve the problem of finding
discrete logarithms over finite groups.  It is this ability that
makes quantum computers capable, in principle, of undermining the
security of elliptic curve cryptographic systems, which are widely
used by industry and government to protect sensitive information.
There is no known classical algorithm for solving the discrete
logarithm problem in polynomial time.  In 1994, Peter Shor
\cite{Sho94} described a quantum algorithm for solving this
problem in polynomial time.

The construction of medium- or large-scale quantum computers has
turned out to be an enormous technological challenge.  For most of
the proposed (practical) schemes for implementing quantum
computers, qubits are a very `expensive' resource.  Thus there is
a significant practical interest in optimizing quantum algorithms
to use as few qubits as possible.  In \cite{PZ03}, Proos and Zalka
give an optimized implementation of the discrete logarithm
algorithm, for the particular case of elliptic curve groups.  They
consider only elliptic curves over the prime fields $\GF(p)$. Many
elliptic curve cryptosystems use elliptic curves over the binary
fields $\GF(2^m)$ however.  So it is important to examine the
number of qubits required to implement the discrete logarithm
algorithm for elliptic curve groups over these binary fields. In
this direction, we show how to decompose the group operation into
a series of smaller, individually reversible, steps (following the
approach taken in \cite{PZ03}). Some of these steps will involve
divisions of elements in the binary field $\GF(2^m)$. To solve
this problem, we show how to implement the extended Euclidean
algorithm for polynomials, and optimize this implementation to use
few qubits.

\section{Elliptic curves over $\GF(2^m)$}

An \emph{elliptic curve} over a field $F$ is the set of points
$(x,y)\in F^2$ satisfying
\[y^2+a_1xy+a_3y=x^3+a_2x^2+a_4x+a_5,\]
subject to some additional conditions on the constants
$a_1,\ldots, a_5\in F$, together with a `point at infinity',
denoted $\O$. For the particular case of curves over the finite
fields $\GF(2^m)$, the defining equation and additional conditions
simplify as follows.
\begin{enumerate}
\item[Case 1:] $a_1\neq 0$ \emph{(non-supersingular curves)}
\[y^2+xy=x^3+ax^2+b\hspace{5mm},\hspace{5mm}b\neq 0.\]

\item[Case 2:] $a_1= 0$ \emph{(supersingular curves)}
\[y^2+cy=x^3+ax+b\hspace{5mm},\hspace{5mm}c\neq 0.\]
\end{enumerate}

An elliptic curve over $\GF(2^m)$ is the set of points $(x,y)\in
\GF(2^m)\times \GF(2^m)$ that satisfy one of the above two formulae,
together with the point at infinity $\O$. A particular curve of one
of the above types is specified by giving values to the constants
$a,b$ (and $c$ in the case of a supersingular curve). The set of
points on a given elliptic curve forms a group under the following
operation of addition.  Let $P=(x_1,y_1)$ and $R=(x_2,y_2)$, where
$P\neq R$, be two distinct points on a curve over $\GF(2^m)$.  The
point $P+R$ is defined as follows.

\begin{enumerate}
\item[Case 1:] \emph{non-supersingular curves}
\[P+R=\begin{cases}\O\hspace{5mm}\text{if }(x_2,y_2)=(x_1,x_1-y_1)\\
(x_3,y_3)\hspace{5mm}\text{otherwise, }\end{cases}\]
\[\text{where}\hspace{5mm}x_3=\lambda^2+\lambda+x_1+x_2+a\hspace{5mm},\hspace{5mm}y_3=\lambda(x_1+x_3)+x_3+y_1\]
\[\lambda=\frac{y_1+y_2}{x_1+x_2}.\]

\item[Case 2:] \emph{supersingular curves}
\[P+R=\begin{cases}\O\hspace{5mm}\text{if }(x_2,y_2)=(x_1,y_1+c)\\
(x_3,y_3)\hspace{5mm}\text{otherwise, }\end{cases}\]
\[\text{where}\hspace{5mm}x_3=\lambda^2+x_1+x_2\hspace{5mm},\hspace{5mm}y_3=\lambda(x_1+x_3)+y_1+c\]
\[\lambda=\frac{y_1+y_2}{x_1+x_2}.\]
\end{enumerate}
Following the argument in \cite{PZ03}, we can avoid dealing with
the cases $P=R$ (point doubling) $P=-R$, and $R=\O$, and restrict
ourselves to the generic group addition formulae in terms of
$x_3,y_3$ above. The key observation is that in a superposition
(such as we would have in the quantum discrete logarithm
algorithm), situations other than the generic case will occur for
only a small fraction of the elements in superposition, and so by
ignoring them the fidelity loss will be negligible.

\section{The discrete logarithm algorithm for elliptic curve groups}

Let $G$ be a cyclic group, and let $\alpha$ be a generator for $G$.
The discrete logarithm problem with respect to the base $\alpha$ is
the following.  Given a group element $\beta\in G$, find the unique
integer $d\in [0,|G|-1]$ such that $\beta=\alpha^d$. Recall that
Shor's quantum algorithm for solving the discrete logarithm problem
makes use of a unitary operator that performs
\[\ket{x}\ket{y}\ket{z}\rightarrow \ket{x}\ket{y}\ket{z\oplus\alpha^x\beta^y},\hspace{1cm}(\ast)\]
where $x$ and $y$ are integers in the range $[0,\ldots,|G|-1]$.

Consider an elliptic curve $E$ and let $P$ be a point on $E$.
Consider the cyclic subgroup of the elliptic curve group generated
by $P$.  We are interested in solving the discrete logarithm
problem for this subgroup.  The group operation is written
additively, so the discrete logarithm problem is the following.
Given a point $Q$ in the subgroup generated by $P$, find the
unique integer $d\in[0,\ldots,\text{order}(P)-1]$ such that
$Q=dP$. The unitary operation $(\ast)$ used in Shor's algorithm
performs
\[\ket{x}\ket{y}\ket{z}\rightarrow \ket{x}\ket{y}\ket{z\oplus (xP+yQ)}.\]
Employing the semiclassical Fourier transform of Griffiths and Niu
\cite{GN95} as detailed in \cite{PZ03}, for the discrete logarithm
algorithm it suffices to be able to implement
\[\ket{S}\rightarrow\ket{S+A}\hspace{5mm}S,A\in E\text{ and
}A\text{ is fixed and `classically known'}.\] Writing $S=(x,y)$
and $A=(\alpha,\beta)$, we want to implement
\[\ket{(x,y)}\rightarrow\ket{(x,y)+(\alpha,\beta)}.\]

\section{Decomposing the group
operation}\label{sec_decompose_group_op}

We now show how to decompose the group operation for curves over
$\GF(2^m)$ into a sequence of individually reversible steps.  Doing
so allows the implementation of the group operation with a smaller
number of ancillary qubits.

We will use the following notation.  When we write $x\rightarrow y$,
we are referring to a (not necessarily reversible) computation
transforming the value $x$ into the value $y$.  When we write $x\iff
y$, we are referring to a \emph{reversible} computation which can be
seen as transforming $x$ into $y$, or as transforming $y$ into $x$.

For a fixed point $(\alpha,\beta)$, define $(x^\prime,y^\prime):=
(x,y)+(\alpha,\beta)$.  We want to decompose the operation
\[\ket{(x,y)}\rightarrow\ket{(x^\prime,y^\prime)}.\]  For simplicity, in
the following we will write the values without the Dirac \emph{ket}
symbols.

\begin{enumerate}
\item[Case 1:] \emph{non-supersingular curves}\newline
We have
\[\lambda=\frac{y+\beta}{x+\alpha}=\frac{x^\prime+y^\prime}{x^\prime+\alpha}.\]
The group operation is decomposed as
\begin{align*}
x,y &\iff x+\alpha,y+\beta\iff
x+\alpha,\lambda=\frac{y+\beta}{x+\alpha}\\
&\iff
x^\prime+\alpha,\lambda=\frac{x^\prime+y^\prime}{x^\prime+\alpha}
\iff x^\prime+\alpha,x^\prime+y^\prime\iff
x^\prime,x^\prime+y^\prime\iff x^\prime,y^\prime.
\end{align*}
The second step in the above decomposition is a division, and the
fourth step is a multiplication, where in each case one of the
operands is uncomputed in the process. All the other steps involve
only additions (and the third step also requires the squaring of
$\lambda$). It turns out that the number of qubits required to
perform the group operation is bounded by the number of qubits
required to perform a division or multiplication where one of the
operands is uncomputed in the process.

\item[Case 2:]
\emph{supersingular curves}\newline We have
\[\lambda=\frac{y+\beta}{x+\alpha}=\frac{y^\prime+c+\beta}{x^\prime+\alpha}.\]
The group operation is decomposed as
\begin{align*}
x,y &\iff x+\alpha,y+\beta\iff
x+\alpha,\lambda=\frac{y+\beta}{x+\alpha}\\
&\iff
x^\prime+\alpha,\lambda=\frac{y^\prime+c+\beta}{x^\prime+\alpha}
\iff x^\prime+\alpha,y^\prime+c+\beta\iff x^\prime,y^\prime.
\end{align*}
As in the non-supersingular case, the second step in the above
decomposition is a division, and the fourth step is a
multiplication, where in each case one of the operands is
uncomputed in the process.  The other steps involve only
additions, and one squaring.  So again the qubit-space requirement
for the group operation is that for a division or multiplication
where one of the operands is uncomputed in the process.
\end{enumerate}
In both the supersingular and non-supersingular case, the qubit
space requirement of the group operation is determined by that of
performing a division or multiplication, where one of the operands
is uncomputed in the process. Such a multiplication can be
achieved by running such a division backwards, so we turn our
attention to implementing divisions of the form $x,y\iff x,y/x$,
using as few qubits as possible. Following \cite{PZ03} the
division is decomposed into the following four reversible steps.
\[x,y\overset{E}{\iff}1/x,y\overset{m}{\iff}1/x,y,y/x\overset{E}{\iff}x,y,y/x
\overset{m}{\iff}x,0,y/x.\] The letters over the arrows are $m$
for standard polynomial multiplication, and $E$ for ``Euclid's
algorithm''. The second $m$ is really a standard polynomial
multiplication run backwards to uncompute $y$.  We know how to
implement standard multiplication in $\GF(2^m)$ using $2m$ qubits
by \cite{BBF03}, so it remains to show how to implement the
extended Euclidean algorithm for polynomials to compute inverses
in $\GF(2^m)$.

\section{The extended Euclidean algorithm for polynomials}

Suppose $A(z)$ and $B(z)$ are two binary polynomials in the variable
$z$, of degrees less than $m$ (i.e. $A,B\in\GF(2^m)$). Suppose $A$
and $B$ are not both 0, and are such that $\deg(A)\leq\deg(B)$. The
\emph{greatest common divisor} of $A$ and $B$, denoted $\gcd(A,B)$,
is the binary polynomial of highest degree that divides both $A$ and
$B$.  The classical Euclidean algorithm for finding $\gcd(A,B)$ is
based on the fact that $\gcd(A,B)=\gcd(B-CA,A)$, for all binary
polynomials $C$.  If we divide $B$ by $A$ (by standard long division
of polynomials), obtaining a quotient polynomial $q(z)$ and a
remainder polynomial $r(z)$ satisfying $B=qA+r$, then
$\deg(r)<\deg(A)$.  By the fact observed above, we have
$\gcd(A,B)=\gcd(r,A)$.  The classical Euclidean algorithm for
polynomials makes this replacement repeatedly until one of the
arguments is 0.  If we set $r_0=A$ and $r_1=B$, the Euclidean
algorithm performs the following sequence of divisions:
\begin{align*}
r_0&=q_1r_1+r_2,&&\hspace{-1cm}0<\deg(r_2)<\deg(r_1)\\
r_1&=q_2r_2+r_3,&&\hspace{-1cm}0<\deg(r_3)<\deg(r_2)\\
&\hspace{2mm}\vdots&&\hspace{.2cm}\vdots\\
r_{m-2}&=q_{m-1}r_{m-1}+r_m,&&\hspace{-1cm}0<\deg(r_m)<\deg(r_{m-1})\\
r_{m-1}&=q_mr_m+0.&&
\end{align*}
The fact above gives us the corresponding sequence of equalities:
\begin{equation*}
\gcd(r_0,r_1)=\gcd(r_1,r_2)=\ldots=\gcd(r_{m-1},r_m)=\gcd(r_m,0).
\end{equation*}
At this point we have the result, since $\gcd(r_m,0)=r_m$. The
algorithm is guaranteed to terminate, since the degree of one of
the arguments strictly decreases in each step.  Moreover, the
algorithm is efficient because the number of iterations is bounded
by the degree of $A$ (which is at most $m$).

Recall that the $\gcd$ of two integers $a,b$ can always be written
as a linear combination of $a$ and $b$ having integral
coefficients.  The same is true for the $\gcd$ of two polynomials
$A,B$.  That is, there exist polynomials $k,k^\prime$ in
$\GF(2^m)$ such that
\[\gcd(A,B)=kA+k^\prime B.\]
The \emph{extended Euclidean algorithm for polynomials} is the
same as the Euclidean algorithm for polynomials except that it
also keeps track of the `coefficient' polynomials $k,k^\prime$
above. It does so through the following recurrences.
\[k_j=\begin{cases}1&\text{ if }j=0\\0&\text{ if
}j=1\\k_{j-2}-q_{j-1}k_{j-1}&\text{ if }j\geq 2\end{cases}\] and
\[{k^\prime}_j=\begin{cases}0&\text{ if }j=0\\1&\text{ if
}j=1\\{k^\prime}_{j-2}-q_{j-1}{k^\prime}_{j-1}&\text{ if }j\geq
2.\end{cases}\] It is not hard to show that for $0\leq j\leq m$ we
have $r_j=k_jr_0+{k^\prime}_jr_1$, where the $r_j$'s are defined
as in the Euclidean algorithm for polynomials, and the $k_j$ and
the ${k^\prime}_j$ are defined by the above recurrences.

For reference, we write the extended Euclidean algorithm for
polynomials in pseudo-code below.  The notation $x\leftarrow y$ is
intended to mean that we assign the value of $y$ to the variable
named $x$.
\newpage
\begin{small}
EXTENDED EUCLIDEAN ALGORITHM FOR POLYNOMIALS
\begin{enumerate}
\item[]$A_0\leftarrow A$\vspace{-1mm}
\item[]$B_0\leftarrow B$\vspace{-1mm}
\item[]$k_0\leftarrow 1$\vspace{-1mm}
\item[]$k\leftarrow 0$\vspace{-1mm}
\item[]${k^\prime}_0\leftarrow 0$\vspace{-1mm}
\item[]$k^\prime\leftarrow 1$\vspace{-1mm}
\item[]$q\leftarrow \floor{\frac{A_0}{B_0}}$\vspace{-1mm}
\item[]$r\leftarrow A_0-qB_0$\vspace{-1mm}
\item[]\textbf{while }$r>0$\textbf{ do }\vspace{-1mm}
\begin{enumerate}
\item[]\emph{temp}$\leftarrow {k^\prime}_0-qk^\prime$\vspace{-1mm}
\item[]${k^\prime}_0\leftarrow k^\prime$\vspace{-1mm}
\item[]$k^\prime\leftarrow$\emph{temp}\vspace{-1mm}
\item[]\emph{temp}$\leftarrow k_0-qk$\vspace{-1mm}
\item[]$k_0\leftarrow k$\vspace{-1mm}
\item[]$k\leftarrow$ \emph{temp}\vspace{-1mm}
\item[]$A_0\leftarrow B_0$\vspace{-1mm}
\item[]$B_0\leftarrow r$\vspace{-1mm}
\item[]$q\leftarrow\floor{\frac{A_0}{B_0}}$\vspace{-1mm}
\item[]$r\leftarrow A_0-qB_0$\vspace{-1mm}
\end{enumerate}
\item[]\textbf{return}$(r,k,k^\prime)$
\end{enumerate}
\end{small}

Inverses in $\GF(2^m)$ can be computed using the extended
Euclidean algorithm for polynomials, as follows.  Suppose $f(z)$
is an irreducible polynomial of degree $m$, and let $C(z)$ be a
binary polynomial of degree $\leq m-1$.  Then $\gcd(C,f)=1$, and
the extended Euclidean algorithm for polynomials finds binary
polynomials $k$ and $k^\prime$ such that $kC+k^\prime f=1$.  But
this means that $kC\equiv 1(\mod f)$, and so $k\equiv C^{-1}(\mod
f)$.  The coefficient $k^\prime$ of $f$ is not needed for the
inversion of $C$, and so we only need to record the coefficient
$k$ of $C$ throughout the algorithm.

\section{Naive Implementation of the extended Euclidean algorithm for polynomials}
\label{sec_naive} We now turn our attention to quantum
implementations of the extended Euclidean algorithm for
polynomials for computing the inverse of an element $C$. Following
$\cite{PZ03}$, our implementations will maintain two ordered pairs
$(a,A)$ and $(b,B)$ ., where $A$ and $B$ record the sequence of
remainders in the Euclidean algorithm for polynomials, and $a$ and
$b$ record the updated coefficient of $C$ for each of the past two
iterations of the algorithm . We call these ordered pairs
\emph{Euclidean pairs}. The algorithm begins with $(a,A)=(1,C)$,
and $(b,B)=(0,f)$ (where $f$ is an irreducible polynomial of
degree $m$). Note that $\deg(C)\leq m-1<m=\deg(f)$. We will always
store the Euclidean pair with the smaller-degree polynomial in the
second co-ordinate first. That is, we store the Euclidean pairs in
the order
\[(a,A)\,(b,B)\]
where $\deg(A)<\deg(B)$.  We then want to perform long division of
$B$ by $A$, obtaining a quotient polynomial $q$ and a remainder
polynomial $r$ satisfying $B=qA-r=qA+r$ (the second equality
follows since the field is binary), where $q$ is the quotient
polynomial of $B/A$, which we denote as $q=\floor{B/A}$. We will
then replace $B$ by $r=B+qA$, and $b$ by $b+qa$.  Since
$\deg(r)<\deg(A)$, after the above replacement we will have to
interchange the Euclidean pairs to maintain the ordering so that
the pair with the smaller-degree polynomial in the second
co-ordinate appears first. So one iteration of the algorithm can
be written as
\[(a,A)\,(b,B),0\longrightarrow (b+qa,B+qA)\,(a,A)\,q\hspace{1cm}\text{where }q=\floor{B/A}.\]
At the beginning of the Euclidean algorithm, we start with
$a=1,b=0,A=C,B=f$, and so $\deg(A)<\deg(B)$ and $\deg(a)>\deg(b)$.
It is easy to see that this condition is preserved in every
iteration of the algorithm.  This implies that we will have
$\floor{\frac{b}{a}}=0$. So we can write
\[q=\floor{\frac{b+qa}{a}}.\]
So while $q$ is computed from the second co-ordinates of the
Euclidean pairs $(a,A),(b,B)$, it can be uncomputed from the first
coordinates of the modified Euclidean pairs $(b+qa,B+qA),(a,A)$.
Thus each iteration of the Euclidean algorithm is individually
reversible, and can be written as
\[(a,A)\,(b,B)\iff (b+qa,B+qA)\,(a,A)\hspace{1cm}\text{where }q=\floor{B/A}.\]
This is decomposed into the following three individually
reversible steps:
\begin{align*}
A\,B,0&\iff A\,B+qA\,q\\
a\,b\,q&\iff a+qb\,b,0\\
&\SWAP
\end{align*}
where ``SWAP'' refers to the operation of switching the two
Euclidean pairs.  Since $\deg(b)<\deg(a+qb)$, the second operation
above is simply the reverse of the first operation.

To perform the division $A,B,0\iff A,B+qA,q$ we can use long
division of the binary polynomial $B$ by $A$.  To implement this
long division, the basic idea is to shift $A$ all the way to the
left (i.e. we shift $A$ left by $m-\deg(A)-1$ bits). Then we start
shifting $A$ to the right one bit at a time, each time
conditionally doing a subtraction. For the binary field $\GF(2^m)$
this is simplified by virtue of the fact that subtraction is the
same as addition, and is achieved by a bitwise $\XOR$ operation.
This bitwise XOR can be implemented quantumly using CNOT gates,
and no ancillary qubits. (Furthermore, these CNOTs could in
principle be performed in parallel, allowing us to do addition in
a single step.)  Note that in our long divisions we are doing more
work than necessary. Often the degree of $B$ will be less than
$m-1$, and so it would not be necessary to shift $A$ all the way
to the left (we could just shift it so the most significant bits
of $A$ and $B$ line-up).  For simplicity, in the naive
implementation we do not take advantage of this fact, but will do
so when we look at an optimized implementation.

\subsection{Implementing some tools}\label{sec_tools}
To implement the long division, there are some subcomponents that
we will need to implement.  We describe implementations of some of
these subcomponents here, optimizing for the number of qubits.

In what follows, we will show how to implement some operation, and
then use that operation \emph{controlled} on the value(s) of some
other qubit(s).  We need to consider whether this can be done
without the requirement for any additional qubits, or an
unreasonable increase in the running time.  Fortunately, by
\cite{BBC+95}, given a gate performing $U$, we can construct a gate
performing a \emph{controlled}-$U$ (that is, $U$ conditioned on a
control qubit being in state $\ket{1}$) with no additional ancillary
qubits, and a small overhead in running time. Using this result
repeatedly, we can implement $U$ conditioned on any desired pattern
of control qubits (e.g. $U$ may be applied only when a three-qubits
control register is in the state $\ket{101}$) with no additional
ancillary qubits, and a small overhead in running time. We will use
this result implicitly in the following.

\begin{figure}[htb]
\begin{center}
\input{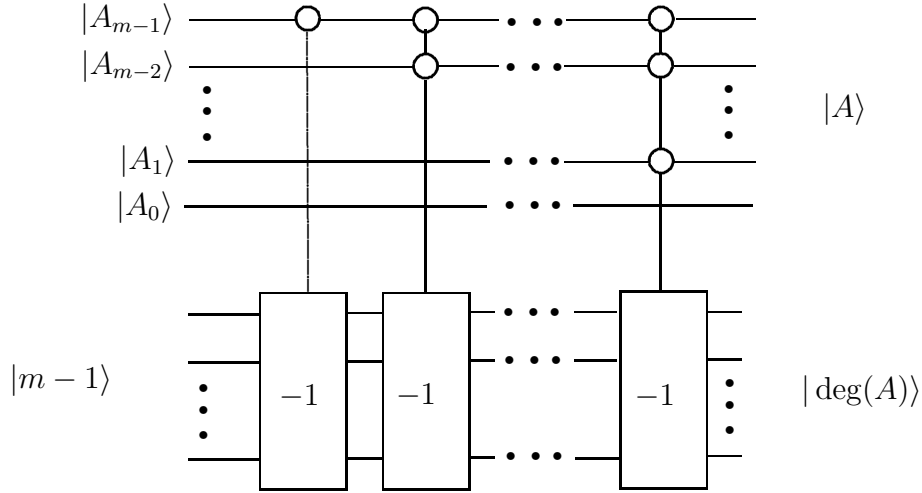}
\end{center}
\caption{\small{Circuit to
compute the degree of $A\in GF(2^m)$.}}\label{fig_degree}
\end{figure}

For the long division, we will need to compute the degree of $A$.
The circuit shown in Figure \ref{fig_degree} accomplishes this.
Each of the hollow circles in the figure denotes a
$0$-\emph{control} (that is, the $(-1)$ operation is applied if
the control qubit is $\ket{0}$). To uncompute the degree, we can
simply run the circuit shown in Figure \ref{fig_degree} backwards.
The circuit in Figure \ref{fig_degree} uses a sequence of $m$
decrementing (-1) gates, each of which is controlled by the values
of some of the qubits of $\ket{A}$.  These decrementing gates update
the value of $\deg(A)$, being computed into a $\ceil{\log
(m-1)}$-qubit register. In Figure \ref{fig_dec}, we show how to
implement an incrementing (+1) gate using only one additional
ancillary qubit.
\begin{figure}[htb]
\begin{center}
\input{dec.pictex}
\end{center}
\caption{\small{Circuit to compute
$\ket{k}\iff\ket{k+1}$.}}\label{fig_dec}
\end{figure}
The ancillary qubit becomes the most-significant-bit of the
result. If we only apply the incrementing circuit to integers in
the range $[0,\ldots,m-2]$, we know that the ancillary qubit will
always be $\ket{0}$ at the output. Decrementing is accomplished by
running this circuit backwards, with the ancillary qubit initially
set to $\ket{0}$. As long as we apply the decrementing circuit to
integers in the range $[1\ldots m-1]$, we know that the ancillary
qubit will always be $\ket{1}$ at the output.  So we can reset the
ancillary qubit to $\ket{0}$ with a NOT gate after each decrement
gate, and re-use that ancillary qubit for the next decrement gate.
Henceforth when we count qubits in this paper, we will always
assume $\ceil{\log (m-1)}=\ceil{\log m}=\ceil{\log (m+1)}$, and
write $\ceil{\log m}$ for convenience.  Similarly for $\floor{\log
m}$. So the degree of $A\in GF(2^m)$ can be computed using
$\ceil{\log m}+1$ qubits (a $\ceil{\log m}$-qubit register into
which the result is computed and stored, and 1 ancillary qubit
shared by the decrementing gates).

\begin{figure}[h]
\begin{center}
\epsfig{file=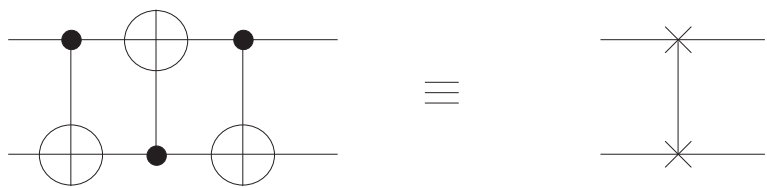}\caption{\small{The quantum SWAP
gate}}\label{fig_swap}
\end{center}
\end{figure}

We also need to implement shifts of our quantum registers.  For
our purpose it will suffice to implement a cyclic shift.  We will
make use of the quantum SWAP gate, which swaps two qubits.  A SWAP
gate can be implemented using 3 CNOT gates, and no ancillary
qubits, as shown in Figure \ref{fig_swap}.  Right shifts can be
implemented by an analogous circuit.

\begin{figure}[h]
\begin{center}
\epsfig{file=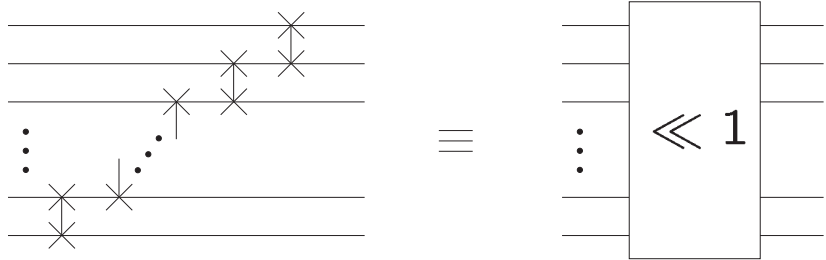}\caption{\small{A cyclic left shift gate
}}\label{fig_shift}
\end{center}
\end{figure}

A left cyclic shift gate which shifts the state of an $n$-qubit
register to the left cyclically by one qubit is implemented using
$n-1$ SWAP gates, and no ancillary qubits, as shown in Figure
\ref{fig_shift}.

\begin{figure}[h]
\begin{center}
\epsfig{file=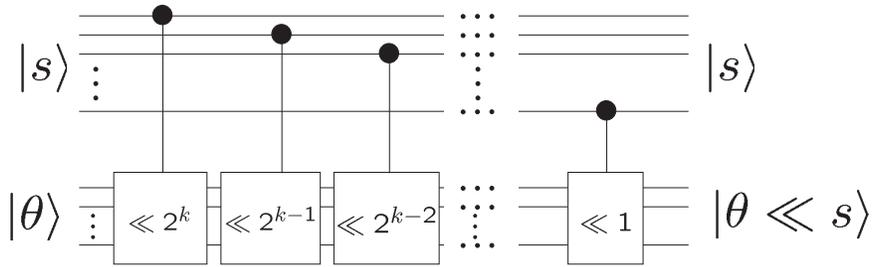}\caption{\small{Circuit for
$\ket{\theta}\ket{s}\iff \ket{\theta<<s}\ket{s}$}.  Here $k=\log_2
s$, and $\ll 2^k$ is implemented by a sequence of $2^k$ $\ll 1$
gates (shown previously).  The overall time complexity is polynomial
in $s$, and no ancillary qubits are required.}\label{fig_cshift}
\end{center}
\end{figure}

A left shift of $s$ qubits can be implemented by concatenating $s$
single-qubit left shifts together. Note that right shifts can be
performed in an analogous manner. We will also need to implement a
shift conditioned on the value contained in a quantum register.
That is, a quantum implementation of the operation
\[\ket{\theta}\ket{s}\iff \ket{\theta<<s}\ket{s}.\]
The controlled shift operation above is implemented by the circuit
shown in Figure \ref{fig_cshift}, where $k$ denotes the number of
bits in the binary representation of $s$.

\subsection{Long division}
Now that we can compute the degrees of polynomials in $\GF(2^m)$,
and perform shifts of quantum registers, we can state an algorithm
to reversibly compute the long division
\[A,B,0\iff A,B+qA,q\] (note the algorithm requires
$\deg(A)<\deg(B)$).

\begin{center}
\begin{boxedminipage}{12cm}
\textbf{Long Division}
\begin{enumerate}
\item[(0)] Initialize $q=0$.
\item[(1)] Compute $\deg(A)$.

\item[(2)] Compute $i=m-\deg(A)-1$.

\item[(3)] Shift $A$ left by $m-\deg(A)-1$ positions.
\item[(4)] While $i\geq 0$ do
\begin{enumerate}
\item[(4.1)] If $B_{i+\deg(A)}=1$, then set $q_{i}=1$ and replace
$B$ with $B \oplus A$.\item[(4.2)] Shift $A$ to the right one bit.
\item[(4.3)] $i\leftarrow i-1$.
\end{enumerate}
\item[(5)] Uncompute $\deg(A)$.
\end{enumerate}
\end{boxedminipage}
\end{center}

At the end of the long division, the register originally containing
$B$ will contain $r=qA+B$. Also, the auxiliary counter $i$ will be
zeroed, and so can be re-used. The conditional setting of $q_i=1$ in
step (4.1) can be accomplished by a CNOT gate, with
$\ket{B_{i+\deg(A)}}$ as the control qubit and $\ket{q_i}$ as the
target qubit. Then, conditioned on $\ket{q_i}$, the operation
$\ket{A,B}\iff \ket{A,A\oplus B}$ can be accomplished by CNOT gates
between the corresponding qubits of $A$ and $B$.  To conditionally
apply this operation, we replace these CNOT gates by Toffoli gates,
with $\ket{q_i}$ as the additional control qubit.

\section{The Problem of Synchronization}

In the discrete logarithm algorithm, the extended Euclidean
algorithm for polynomials will be applied to a superposition of
inputs. For this reason we have to be careful that the steps of the
algorithm are appropriately synchronized, so that each element in
the superposition is undergoing the same step at any given time.  In
the naive implementation described above, we shift $A$ left by
$m-\deg(A)-1$ bits.  The number of computational steps to perform
this shift depends on $\deg(A)$. When the computation is applied to
a superposition of inputs, $\deg(A)$ will be different for the
different elements in the superposition. Thus the number
computational steps is different for different elements in
superposition. This means the stages of the algorithm will not be
properly synchronized between elements in superposition.

This synchronization problem can be solved by applying a general
technique of \emph{synchronizing} the implementation
\cite{PZ03}\footnote{In \cite{PZ03} they refer to the technique as
``\emph{de}synchronization'', but we feel ``synchronizing'' is more
clear.} . We explain synchronization by way of an example.
 Suppose a computation $C$ consists of some sequence of
three simple reversible operations $o_1$, $o_2$ and $o_3$ (and no
other operations). The time taken to perform each of the
operations $o_1,o_2,o_3$ is independent of the input. This means
that on a superposition of inputs, the time required to perform
the operation $o_1$ (for example) is the same for all elements in
the superposition.

\begin{figure}[h]
\begin{center}
\epsfig{file=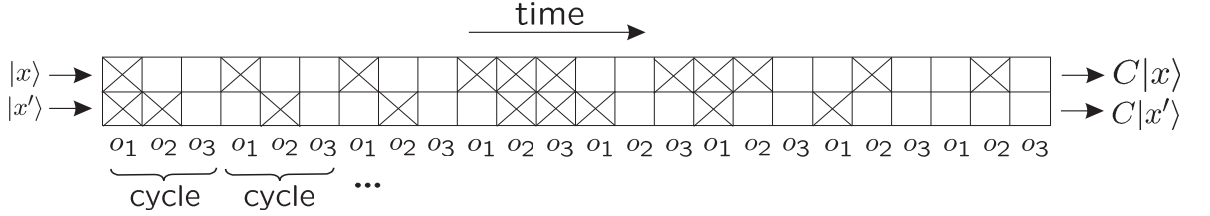}\caption{\small{synchronization
example.}}\label{fig_desynch}
\end{center}
\end{figure}

The quantum computation $C$ is some sequence of the operations
$o_1$, $o_2$ and $o_3$, in any order, and with repetitions.  For
example, $C$ applied to the input basis state $\ket{x}$ might
consist of $o_1$ applied 4 times, followed by $o_2$ applied 1
time, followed by $o_3$ applied 2 times, followed by $o_1$ applied
1 time, followed by $o_2$ applied 3 times. That is,
\[
C\ket{x}=o_2o_2o_2\hspace{2mm}o_1\hspace{2mm}o_3o_3\hspace{2mm}o_2\hspace{2mm}o_1o_1o_1o_1\ket{x}.
\]
The synchronization problem is that for another input basis state
$\ket{x^\prime}$ (in a superposition of inputs), the sequence of
operations might be different.  For example, on $\ket{x^\prime}$
the same computation $C$ might consist of $o_1$ applied 1 time,
followed by $o_2$ applied 4 times, followed by $o_3$ applied 1
time, followed by $o_1$ applied 3 times.  That is,
\[
C\ket{x^\prime}=o_1o_1o_1\hspace{2mm}o_3\hspace{2mm}o_2o_2o_2o_2\hspace{2mm}o_1\ket{x^\prime}.
\]
The idea of synchronization is to have \emph{all} the computations
in the superposition cycle through the 3 operations repeatedly, each
time allowing the computation to either apply the operation once, or
not apply it (wait for the next operation). The cycle is repeated a
sufficient number of times so that sufficiently many of the
computations in superposition have finished.  For the computation
$C$ above applied to the two input basis states $\ket{x}$ and
$\ket{x^\prime}$, this is illustrated in Figure \ref{fig_desynch}.
In the figure, the operation applied at each step are indicated by
an $\times$ in the corresponding box.
We now describe more explicitly how to implement synchronization.
There must be a way for the computation to tell when a series of
$o_i$'s is finished and the next one should begin.  We want to do
this reversibly, so there must be a way to tell both when an $o_i$
is the first in a series, and when it is last in a series. In each
$o_i$ we can include a a sequence of gates which flips a flag qubit
$f$ if $o_i$ is the first in a sequence, and another mechanism that
flips $f$ if $o_i$ is the last in a sequence.  We also make use of a
small ``counter'' register $c$ to control which operation is
scheduled to be applied at the current step. Thus we have a triple
$x,f,c$ where $x$ stands for the actual data.  We initialize both
$f$ and $c$ to 1 to signify that the first operation will be the
first in a sequence of $o_1$ operations. The physical quantum-gate
sequence which we apply is
\[\ldots\hspace{2mm}ac\hspace{2mm}o_1^\prime\hspace{2mm}ac\hspace{2mm}o_3^\prime\hspace{2mm}
ac\hspace{3mm}o_2^\prime\hspace{2mm}ac\hspace{2mm}o_1^\prime\hspace{2mm}
ac\hspace{2mm}o_3^\prime\hspace{2mm}ac\hspace{2mm}o_2^\prime\hspace{2mm}
ac\hspace{2mm}o_1^\prime\hspace{2mm}\ket{x}\] where the
$o_i^\prime$ are the $o_i$ conditioned on $i=c$ and $ac$ stands
for ``advance counter''.  These operations act as follows on the
triple:
\begin{align*}
&o_i^\prime:\hspace{3mm}\text{if
}i=c:\hspace{2mm}x,f,c\hspace{1mm}\leftrightarrow\hspace{1mm}
o_i(x),f\oplus\text{first}\oplus\text{last},c\\
&ac:\hspace{3mm}x,f,c\hspace{1mm}\leftrightarrow\hspace{1mm}
x,f,(c+f)\mod 3
\end{align*}
where $o_i^\prime$ does nothing if $i\neq c$, the symbol
``$\oplus$'' means XOR, and $(c+f)\mod 3$ is taken from
$\{1,2,3\}$.  In the middle of a sequence of $o_i$'s the flag $f$
is 0, and so the counter doesn't advance.  The last in a sequence
of $o_i$'s will set $f=1$ and the counter will advance in the next
$ac$ step.  The first operation of the next series resets $f$ to
0, so that this series can progress.

Of course, even though the individual steps in the algorithm are
synchronized, the computations in the superposition will in
general finish the extended Euclidean algorithm after different
numbers of iterations.  For those that finish earlier than others,
we cannot simply have them ``halt'' and wait for the others to
finish (this would result in an implementation that is not
reversible).  To ensure reversibility, those elements in
superposition that halt early must increment a small counter at
each time step until the other elements in superposition finish.
We will call this small counter the ``halting counter''.

We do not describe in detail how to apply synchronization to repair
the naive implementation, but instead proceed with a better
optimized implementation that will make use of synchronization.

\section{An optimized implementation}\label{sec_opt}
\subsection{The implementation}

The starting point for an optimized implementation is the
observation that large quotients occur relatively rarely in the
extended Euclidean algorithm for polynomials.  In the naive
implementation by shifting $A$ all the way to the left in the long
divisions, we were doing more work than necessary.  Our optimized
implementation will make use of ``adaptive'' long divisions, whose
behaviour is conditioned on the sizes of the arguments.  In fact,
any $O(n^2)$ algorithm (classical or quantum) must do this kind of
adaptive division.  For a quantum implementation, we will then
note that since large quotients occur rarely, we can bound the
size of the quotient with a negligible loss in fidelity.

The other main observation underlying the optimized implementation
is that in the naive implementation we were using much more space
than necessary to store the Euclidean pairs.  In the naive
implementation we used a separate $m$-qubit register for each of
$A,B,a,b$.  It turns out that this is twice as much space as is
necessary.
\begin{claim}
At every stage of the extended Euclidean algorithm for polynomials
we have $\deg(aB)=m$.
\end{claim}
\begin{quote}
\proof Initially we have $aB=f$ and so $\deg(aB)=m$, so the claim is
true at the first iteration.  Each iteration transforms
\begin{align*}
a&\rightarrow a^\prime=b+qa\\
B&\rightarrow B^\prime=A.
\end{align*}
So we have
\begin{align*}
\deg(a^\prime B^\prime) &= \deg((b+qa)A)\\
&=\deg(qaA)\hspace{5mm}\text{(since $\deg(qa)\geq\deg(a)>\deg(b)$)}\\
&=\deg(q)+\deg(a)+\deg(A)\\
&=\deg(B)-\deg(A)+\deg(a)+\deg(A)\\
&=\deg(aB)\\
&=m
\end{align*}
and so the claim is true after each iteration.$\qed$
\end{quote}

An immediate corollary of this claim is
\begin{cor}
At every stage of the extended Euclidean algorithm for polynomials
we have
\[\deg(a)+\deg(A)\leq
m\hspace{5mm}\textnormal{and}\hspace{5mm}\deg(b)+\deg(B)\leq m.\]
\end{cor}
\begin{quote}
\proof Since $\deg(A)<\deg(B)$ we have
\[\deg(a)+\deg(A)=\deg(aA)\leq\deg(aB)=m.\]
Similarly, since $\deg(a)>\deg(b)$ we have
\[\deg(b)+\deg(B)=\deg(bB)\leq\deg(aB)=m.\qed\]
\end{quote}
By the corollary, we see that a single $m$-qubit register will be
sufficient to store both $a$ and $A$, and a second $m$-qubit
register is sufficient to store both $b$ and $B$.  Thus $A$ and
$a$ can \emph{share} a single $m$-qubit register, and $b$ and $B$
can share a second $m$-qubit register.  This reduces the total
space to store $A,B,a,b$ from $4m$ to $2m$. The problem with this
approach is that the relative sizes of $a$ and $A$ change from one
iteration to the next, and thus so does the boundary between $A$
and $a$ within the single $m$-qubit register (similarly for $b$
and $B$).  Further, at any iteration, this boundary may be
different between elements in superposition. So we need a way to
quantumly calculate the position of this boundary for each
iteration.

\begin{figure}[h]
\begin{center}
\epsfig{file=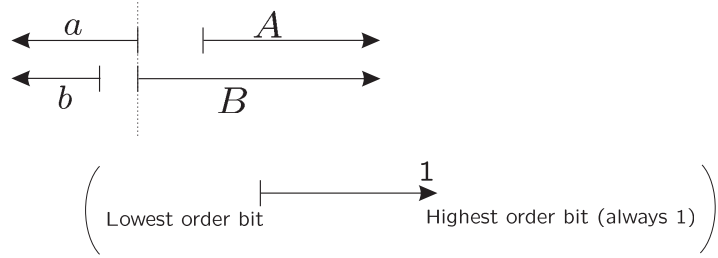}\caption{\small{The positions of
$A,B,a,b$ for register sharing.}}\label{fig_reg_share}
\end{center}
\end{figure}

First, observe that the boundary between $A$ and $a$ can be at the
same position as the boundary between $B$ and $b$, in any
iteration (since $\deg(A)<\deg(B)$).  Second, notice that the
boundary can be easily determined if we know the degrees of
$A,B,a,b$. It will turn out to be convenient to store $A$ and $a$
in a single register in opposing directions.  That is, the most
significant bit of $A$ is at one end of the register, and the most
significant bit of $a$ is at the extreme other end of the
register.  Between $A$ and $a$ the register will be padded with
zeros.  Similarly for $B$ and $b$. The situation for register
sharing is illustrated in Figure \ref{fig_reg_share}.

From Figure \ref{fig_reg_share} it can be seen that the boundary
for register-sharing can be determined from $\deg(a)$ or from
$\deg(B)$.  Our strategy will be to store the degree of each of
$A,B,a,b$ at each step, and use either $\deg(a)$ or $\deg(B)$
(depending on what operation we are performing) to determine the
boundary.  For convenience, we will keep track of the degrees of
all of $A,B,a$ and $b$, requiring 4 separate $\ceil{\log m}$-qubit
registers.

As before, we focus on implementing the long division
\[A,B,0\hspace{1mm}\leftrightarrow\hspace{1mm}A,B+qA,q.\]
The long division algorithm is modified slightly as a result of the
new strategy for storing $A$ and $B$.  Note that we do not need to
initially shift $A$ all the way towards the high order end, since
the most significant bits of $A$ and $B$ are already in the same
position. Instead of shifting $A$ one bit at a time towards the low
order end at each step, we shift $B$ one bit at a time towards the
high order end.  At each stage, a new bit of $q$ is first read out
from the high order bit of $B$.  Then, controlled on the new bit of
$q$ (equivalently the high order bit of $B$) $B$ is XORed with $A$
(this is the conditional subtraction).  Then $B$ is shifted towards
the high order end by 1 bit, and the value of $\deg(B)$ is
decremented by 1.  Note that no significant bits of $B$ are lost in
the shift, because after the conditional XOR operation, we know the
high order bit of $B$ will be 0. After the long division is
complete, the remaining operation is to shift off any leading (high
order) zeros in the final value of $B$, and decrement the value of
$\deg(B)$ accordingly.  This is done so that the most significant
bits of $A$ and $B$ are in corresponding positions for the next
iteration. The operations $o_1$ and $o_2$ for implementing the long
division in a synchronized manner are as follows:
\begin{enumerate}
\item[$o_1$:]

\begin{enumerate}
\item[$(a)$] The high-order bit of $B$ becomes the next bit of $q$
(starting at the high-order bit of $q$ and working down).
\item[$(b)$] Conditioned on the new bit of $q$, $B$ is replaced with
$B\oplus A$. \item[$(c)$] $B$ is shifted towards the high order
end by 1 bit, and $\deg(B)$ is decremented by 1.
\end{enumerate}
\item[$o_2$:] $B$ is shifted towards the high order end by 1 bit,
and $\deg(B)$ is decremented by 1.
\end{enumerate}
The first in a sequence of $o_1$ operations is recognized by the
condition $q=0$.  The last in a sequence of $o_1$ operations is
recognized by $\deg(A)=\deg(B)$.  When performing the last in a
sequence of $o_1$ operations, only part $(a)$ is performed (so
parts $(b)$ and $(c)$ can be conditioned on the flag qubit). The
first in a sequence of $o_2$ operations is recognized by
$\deg(A)=\deg(B)$. The last in a sequence of $o_2$ operations is
recognized when the bit in the high-order ``slot'' of the register
containing $B$ is is $\ket{1}$.

\begin{figure}[h]
\begin{center}
\epsfig{file=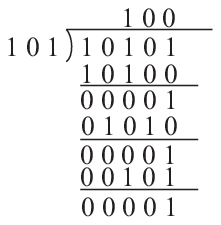}\caption{\small{Example long division by
hand.}}\label{fig_longdiv}
\end{center}
\end{figure}

\begin{figure}
\begin{center}
\epsfig{file=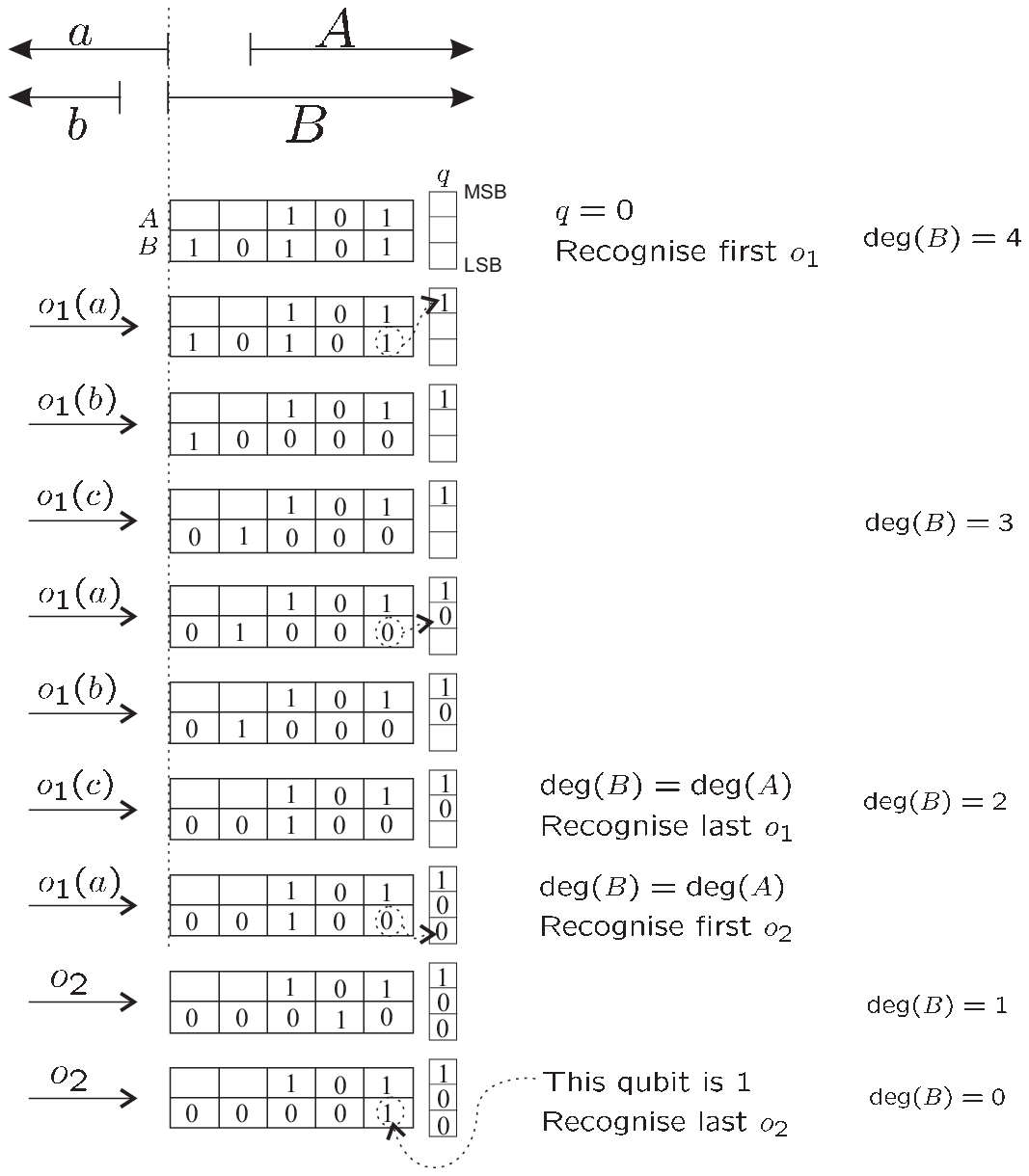}\caption{\small{Example of optimized
implementation of long division.}}\label{fig_example}
\end{center}
\end{figure}

The long division algorithm is illustrated by an example. Suppose
we have the following:
\begin{align*}
A&=z^2+1 \hspace{1.5cm}(A=101)\\
B&=z^4+z^2+1\hspace{.7cm}(B=10101).
\end{align*}
The long division $B/A$ as would be performed by hand is shown in
Figure \ref{fig_longdiv}.
The long division as performed by the algorithm is shown in Figure
\ref{fig_example}.  One feature of the algorithm suggested by the
example is that the qubits can be spatially arranged so that
operations are performed on neighbouring qubits.  Note that in the
implementation of shifts (Figure \ref{fig_shift}) the CNOT gates
are between adjacent qubits as well). This might be advantageous
for a given physical implementation. In Figure \ref{fig_example},
note that blank cells contain the value 0, but are shown as blank
to make it easier to understand the steps of the long division.

We have omitted the details of how to condition the steps of the
long division on the value which determines the boundary for
register sharing.  For example, in the implementation of
$A,B,0\leftrightarrow A,B+qA,q$, the operations on $A,B,q$ will be
conditioned on the value in the register containing $\deg(a)$ (from
which the boundary position for register sharing can be determined).
These details are very complicated, but the techniques for
implementing controlled-gates in $\cite{BBC+95}$ indicate that it
can be done with no ancillary qubits, and a polynomial increase in
time.

\subsection{Qubit space complexity}
We saw in Section \ref{sec_decompose_group_op} that the number of
qubits required to implement the elliptic curve group operation is
bounded by the number of qubits required to implement the extended
Euclidean algorithm for polynomials.  Here we count the number of
qubits required by our implementation.

By using register sharing, the values of $A,B,a,b$ can be stored
using $2m$ qubits.  The values of $\deg(A),\deg(B),\deg(a),\deg(b)$
must be initially computed and stored, requiring $4\ceil{\log{m}}+4$
qubits (as seen in Section \ref{sec_tools}).  We also need to store
the value of the quotient $q$.  We noted that in the extended
Eulcidean algorithm for polynomials large quotients are rare.  In
\cite{PZ03} it is shown that by bounding the size of $q$ to
$3\ceil{\log m}$ bits, the total loss of fidelity will be at most
$\frac{12}{m}$, which is acceptable in the context of Shor's
algorithm.  So we store $q$ in a register of $3\ceil{\log m}$
qubits.

For the synchronization we need a flag qubit $f$, and 2-qubit
counter register $c$ (to index the 4 operations $o_1(a), o_1(b),
o_1(c),$ and $o_2$ used in the synchronization).  Recall that we
also need a ``halting counter'', as the computations in the
superposition will finish the extended Euclidean algorithm for
polynomials after different numbers of iterations.  The exact size
of this halting counter depends on the exact time complexity of the
algorithm.  However, as our implementation is clearly polynomial in
$m$, we know that the size of the halting counter will be at most
logarithmic in $m$.  We will write $H$ for the number of qubits
required for the halting counter, where it is understood that $H$ is
$O(\log m)$.  Such a halting counter would be required in any
quantum implementation of the extended Euclidean algorithm for
Polynomials.

So we have that the qubit space complexity for our implementation of
the extended Euclidean algorithm for polynomials, and thus of the
elliptic curve group operation for curves over $\GF(2^m)$, is
\begin{align*}
&\underset{A,B,a,b}{\underbrace{2m}}+\underset{q}{\underbrace{3\ceil{\log
m}}}
+\underset{\deg{A},\deg{B},\deg{a},\deg{b}}{\underbrace{4\ceil{\log{m}}+4}}+\underset{f,c}{\underbrace{1+2}}+H\\
&=2m+7\ceil{\log{m}}+7+H.
\end{align*}

\section*{Acknowledgements}
This research was supported by MITACS (Mathematics of Information
Technology and Complex Systems), NSERC (National Science and
Engineering Research Council), CSE (Communications Security
Establishment), CSI (Canadian Foundation for Innovation), ORCDF
(Ontario Research and Development Challenge Fund), and PREA
(Premier's Research Excellence Awards).


\begin{thebibliography}{99}

\let\bib\bibitem
    \bib[BBC+95]{BBC+95}
A. Barenco, C.H. Bennett, R. Cleve, D.P. DiVincenzo, N. Margolus,
P. Shor, T. Sleator, J. Smolin, H. Weinfurter, ``Elementary gates
for quantum computation'', Phys. Rev. A, 52:3457-3476, 1995.

\let\bib\bibitem
    \bib[BBF03]{BBF03}
Stephane Beauregard, Gilles Brassard, Jose Manuel Fernandez,
``Quantum Arithmetic on Galois Fields'', Quant-ph/0301163.

\let\bib\bibitem
        \bib[GN95]{GN95}
R.B. Griffiths, C.S. Niu, ``Semiclassical Fourier Transform for
Quantum Computation'', \emph{Phys. Rev. Lett.} 76 (1996) pp.
3228-3231.

\let\bib\bibitem
        \bib[HMV]{HMV}
D. Hankerson, A. Menezes and S. Vanstone, ``Guide to Elliptic
Curve Cryptography'', Springer-Verlag, 2003.

\let\bib\bibitem
    \bib[Sho94]{Sho94}
Peter Shor,  ``Algorithms for Quantum Computation: Discrete
Logarithms and Factoring'', \emph{Proceedings of the 35th Annual
Symposium on Foundations of Computer Science} (1994), 124-134.

\let\bib\bibitem
        \bib[VBE95]{VBE95} V. Vedral, A. Barenco, A. Ekert,
        ``Quantum networks for elementary arithmetic operations'',
        \emph{Phys. Rev. A}, \textbf{54}, 147.

\let\bib\bibitem
        \bib[PZ03]{PZ03}
Christof Zalka, John Proos. ``Shor's discrete logarithm quantum
algorithm for elliptic curves'', QIC Vol. 3 No. 4, pp 317-344
(2003), also quant-ph/0301141.

\end{thebibliography}
\end{document}